\documentclass[12pt,preprint]{aastex}

\usepackage{graphicx}

\slugcomment{}

\title{AGN Obscuring Tori Supported by Infrared Radiation Pressure}

\shortauthors{Krolik}

\begin{document}
\newcommand{\rnorm}{{r \over r_{\rm in}}}
\newcommand{\znorm}{{z \over r_{\rm in}}}

\shorttitle{}

\author{Julian H. Krolik}
\affil{Department of Physics and Astronomy, Johns Hopkins University, 
    Baltimore, MD 21218}

\begin{abstract}

    Explicit 2-d axisymmetric solutions are found to the hydrostatic
equilibrium, energy balance, and photon diffusion equations within
obscuring tori around active galactic nuclei.  These
solutions demonstrate that infrared radiation pressure can support
geometrically thick structures in AGN environments subject to
certain constraints: the bolometric luminosity must be roughly
$\sim 0.03$--$1 \times$ the Eddington luminosity; and the Compton
optical depth of matter in the equatorial plane should be $\sim 1$,
with a tolerance of about an
order of magnitude up or down.  Both of these constraints are at least
roughly consistent with observations.   In addition, angular momentum must be
redistributed so that the fractional rotational support against gravity
rises from the inner edge of the torus to the outer in a manner specific
to the detailed shape of the gravitational potential.  This model
also predicts that the column densities observed in obscured AGN should range
from $\sim 10^{22}$ to $\sim 10^{24}$~cm$^{-2}$.

\end{abstract}


\section{Background}

     Geometrically and optically thick belts of matter can be seen around
many active galactic nuclei (Antonucci 1993).  Although the evidence is best
for nearby Seyfert galaxies (e.g., the direct detection by Jaffe et~al. 2004
of the torus in NGC~1068), there is also strong indirect evidence that this picture
applies to radio galaxies (Barthel 1989, di Serego Alighieri et~al. 1994) and
quasars (e.g., the spectropolarimetry presented in
Zakamska et~al. 2005).  The ratio of observed numbers of obscured AGN to
unobscured gives a measure of the fraction of opaque solid angle; it is
generally $\sim 1$ (Hao et~al. 2005).   X-ray column densities to obscured nuclei
range from $\sim 10^{22}$~cm$^{-2}$ to in excess of $10^{24}$~cm$^{-2}$
(Risaliti et~al. 1999, Treister et~al. 2004); applying a Galactic
dust/gas ratio would imply a dust extinction $A_V \sim 10$---$10^3$~mag,
easily ample for blocking the entire optical and ultraviolet continuum.

    One of the principal puzzles about these tori is how they stand up
against gravity.  Orbital speeds in galactic potentials are generically
$\sim 100$~km~s$^{-1}$, and the nearby black hole can only increase this
characteristic speed.  To stretch upward enough to block a significant fraction
of solid angle requires a vertical velocity component that is a sizable
fraction of the orbital speed; if the velocities are interpreted as thermal,
temperatures of at least $\sim 10^5$~K are implied.  If the gas is this hot,
how can dust survive?

    Numerous ideas have been proposed to answer this question, but none has
proved entirely satisfactory.  The first suggestion (made by Krolik \& Begelman
1988 and elaborated by Beckert \& Duschl 2004) was that the gas and dust are
highly clumped, and the clumps have highly supersonic motions.  To support
these motions against the inevitable losses in cloud-cloud collisions, orbital
shear energy can be randomized if magnetic fields make
the clouds sufficiently elastic.  The central problem with this scheme is that,
although physically possible, the required magnetic
field strengths are not terribly plausible.  Another idea, almost as old,
is that the gas and dust
are always {\it locally} geometrically thin, but the plane in which they
orbit varies as a function of radius (Sanders et~al. 1989).  Countering
this suggestion are both the indirect evidence of well-formed ``ionization
cones" and polarized reflection regions collimated not far from
the innermost part of the torus, and the direct evidence provided by
detection of a geometrically
thick structure right at the torus's inner edge.  Once again, one obtains
the clearest view of all three sorts of data from NGC~1068 (Capetti
et~al. 1997, Kishimoto 1999, Jaffe et~al. 2004).  Still another suggestion
was made by K\"onigl \& Kartje (1994), who argued that the large vertical
motions required by geometrical thickness are best explained as arising
in a magneto-centrifugal wind.  The principal drawbacks to this idea are
the unknown origin of the large-scale magnetic field and the large energy
source needed to drive the wind.  Still another idea is that magnetic fields
alone support a static equilibrium (Lovelace et~al. 1998).
The last notional explanation to list
is that the large optical/ultraviolet radiation flux
of the AGN is converted to mid-infrared by dust at the inner edge of the
torus, and the large opacity of dust in that band couples the radiation
so strongly to the torus matter that the
radiation force is comparable to gravity (Pier \& Krolik 1992a).  When first
proposed, this scheme was only an order of magnitude estimate because
its authors did not attempt to find a self-consistent solution of
both the infrared transfer problem and the force balance problem.

  It is the object of this paper to show, via an idealized model, that self-consistent
equilibrium solutions in which the torus is supported by infrared radiation
force do exist.  The model presented is admittedly
highly simplified and rests on a number of rough approximations.  However,
it does contain all the zeroth-order physics of the problem, and, as
will be shown, it is completely solvable analytically.  Moreover, the
solutions found demonstrate certain qualitative features that should
apply to any more realistic description.

\section{The Model}

\subsection{Qualitative presentation}

      Obscuring tori wrap around a central active nucleus.  In their central
``hole", the matter must be largely transparent, so that viewers located along
the axis can see the nucleus clearly (they then see a ``type 1" AGN).  On
the other hand, on oblique and equatorial lines of sight, there is so much
opacity to wavelengths from the near-infrared to soft X-rays that
viewers in those directions see nothing but infrared radiation from the
warm dust in the torus or very hard X-rays.  The very fact that these
tori are, indeed, ``obscuring" means that they must act as light
reprocessing machines which are heated from the inside and cooled from the
outside.  Radiation flux from the AGN passes through the torus (although
suffering some drastic changes in spectrum {\it en route}), entering
through its inner edge and departing through its upper and outer surfaces.

      The key point behind the idea of supporting the geometric thickness
of the torus by infrared radiation pressure is that the acceleration due to a given
radiation flux $\vec{\cal F}$ is $\kappa \vec{\cal F}/c$, where $\kappa$ is
the opacity per unit mass.  The luminosity $L_{E,eff}$ capable of balancing the
gravity of a mass $M$ in spherical symmetry (the ``effective Eddington luminosity")
is therefore $4\pi c GM/\kappa$, inversely proportional to the opacity.  Because the
opacity of dust per unit mass of gas is an order of magnitude greater than
the Thomson opacity per unit mass when the radiation temperature is
in the range 100--1000~K (Semenov et~al. 2003), $L_{E,eff} \sim 0.1 L_E$
for predominantly infrared light passing through dusty gas (this argument
can be rephrased more precisely in terms of the divergence of the
radiation pressure tensor, but its essential grounding still lies in the relatively
high opacity of dusty gas).  Enhanced
heavy-element abundances likely lead to larger than local dust/gas ratios;
where that is the case, $L_{E,eff}/L_E$ might be even smaller.  Luminosities
$\sim 0.1 L_E$ are commonly expected in AGN, so if a significant part of
the flux striking the inner surface of the torus can be converted to mid-IR
wavelengths, radiation forces comparable to gravity can easily result.

      Rotational support against gravity is so common in astrophysical contexts
that it is entirely plausible to suppose that it is important here, too.  However,
the matter of the torus must have an effective collision rate at least as
large as the orbital frequency.  If it is fluid, this follows by definition.  If
it is highly clumped, this condition is compelled by the requirement that the torus
be consistently opaque (Krolik \& Begelman 1988).  When the average line of
sight through the torus has at least one clump on it, the mean collision rate
of clumps is at least the orbital frequency.  If these collisions are at
all dissipative, one might expect the torus matter to settle into the plane
normal to the total angular momentum and no longer be geometrically thick.

      Adding radiation to the picture changes this conclusion.
Radiation diffusing through the equatorial plane of a geometrically thin but
optically thick annular structure quickly develops a large vertical flux
in the region just outside its inner edge because that is the most direct path
out of the opaque matter.  At the order of magnitude level, this vertical flux is
$\sim (H/r_{\rm in})L/(\pi r_{\rm in}^2)$, where the inner radius of the annulus
is $r_{\rm in}$ and its half-thickness is $H$.  The upward acceleration it
creates, $(\kappa/c)(H/r_{\rm in})L/(\pi r_{\rm in}^2)$, competes with the
downward acceleration of gravity, $(H/r_{\rm in})(GM/r_{\rm in}^2)$.  Because
the same factor of $H/r_{\rm in}$ enters the expressions for both the
acceleration due to radiation and the acceleration due to gravity, if the magnitude
of the flux is comparable to or greater than the effective Eddington flux,
the annulus expands vertically.  Greater thickness leads, of course, to
interception of more light in a manner exactly balancing the growing
magnitude of the vertical gravity.

      A corollary of oblique flux producing a force comparable to gravity
is that the radial component of the radiation force is also comparable to gravity.
If this is so, radial force balance demands a rotation rate that is sub-Keplerian.
Because the radial flux diminishes outward faster than $\propto r^{-2}$ as flux
is diverted to the vertical direction, in equilibrium the specific angular momentum
of the torus must increase outward, approaching Keplerian.

\subsection{Self-consistent solution in the torus interior}

      In a real obscuring torus, the optical through soft X-ray continuum of
the active nucleus is absorbed in a thin layer along its inner edge, where
warm dust reprocesses the nuclear luminosity into the infrared.  This inner
edge is a complicated place, as we will discuss below.  We therefore begin
with the simpler problem of finding a self-consistent description of dynamics
and radiation transfer in the torus interior.  A more precise definition of
the boundaries of this region, and therefore the domain of applicability
of these results, will be given in the following subsection.

      The simplest non-trivial geometry in which this picture can be explored
is 2-d axi\-symmetry.  Adopting cylindrical coordinates $r$ and $z$, we write
the equation of hydrostatic equilibrium for radiation and rotation balancing
gravity as
\begin{equation}\label{eq:hydrostat}
\kappa\vec{\cal F}/c = -\vec g_{\rm eff} = 
   r\Omega^2 (1-j^2)\hat r + z\Omega^2 \hat z,
\end{equation}
where $\Omega$ is the local orbital frequency and the gas's specific angular
momentum is $jr^2\Omega$.   Note that we are supposing that gas pressure
gradients are entirely negligible.  We also adopt three simplifying assumptions, all
appropriate to flattened geometries.  First, we take $\Omega$ at all heights $z$
to be the rotation rate of a circular orbit in the torus midplane at cylindrical
radius $r$.  Second,
we follow only the component of angular momentum parallel to the torus axis, and
$j$ is assumed to be a function of $r$ alone.   Third, we approximate the vertical
component of the gravity by $z\Omega^2$.   This last approximation would be exact
if we evaluated $\Omega$ at the actual local value of $z$, rather than at $z=0$.

   Ideally, to find the flux, one would solve a complete transfer problem at all
relevant frequencies for all photon directions.  Here we take a much simpler
approach: the gray diffusion approximation, using a thermally-averaged opacity.
In this approximation, the flux is given by
\begin{equation}
\vec{\cal F} = -\frac{c}{3\kappa\rho} \nabla E,
\end{equation}
where $\rho$ is the gas density and $E$ is the radiation energy density.  Under the
assumption of hydrostatic balance (eqn.~\ref{eq:hydrostat}), the radiation energy
density and the dynamics are related by
\begin{equation}\label{eq:eraddyn}
-\frac{1}{3\rho} \nabla E = r\Omega^2 (1-j^2)\hat r + z\Omega^2 \hat z.
\end{equation}

If the only source of infrared radiation is the conversion via dust reradiation
of optical and ultraviolet photons at the inner edge of the torus, then in the
body of the torus
\begin{equation}\label{eq:fluxcons}
\nabla \cdot \vec{\cal F} = 0.
\end{equation}
It can also be of interest to explore the effect of distributed sources of
infrared photons.  These may be created, for example, by local heating due to
Compton recoil when hard X-rays penetrate deep in the torus material (as
discussed, e.g., in Chang et~al. 2006) or by the absorption of locally-generated
starlight.  When there are distributed sources, the right-hand-side of
equation~\ref{eq:fluxcons} may be non-zero.  For the purposes of this paper,
however, we adopt the simple assumption of no internal sources.  With that
assumption, the diffusion equation becomes
\begin{equation}\label{eq:fluxconsdiff}
\nabla \cdot \frac{c}{3\kappa\rho}\nabla E = 0.
\end{equation}
Combining equation~\ref{eq:fluxconsdiff} with equation~\ref{eq:eraddyn} gives
\begin{equation}\label{eq:fluxhydrostat}
\nabla \cdot \left(\frac{c}{\kappa}\vec g_{\rm eff}\right) =
     \nabla \cdot \left\{\frac{c}{\kappa}\left[r\Omega^2 (1-j^2)\hat r + 
                    z\Omega^2 \hat z\right]\right\} = 0.
\end{equation}

     Detailed radiation transfer studies of obscuring tori consistently find that
their interior temperatures are in the range 100--1000~K (Pier \& Krolik 1992b,
Efstathiou \& Rowan-Robinson 1995, Granato et ~al. 1997, Nenkova
et~al. 2002).  According to the most recent dust opacity models (e.g.,
Semenov et~al. 2003), the Rosseland mean opacity for gas of Solar abundances
and normal dust content is $\simeq 10$--30 times
greater than Thomson and has no consistent trend within
this temperature range, instead exhibiting only a few mild local maxima and
minima; the ratio of the largest opacity to the smallest is no more than
$\sim 3$.  On this ground, we approximate $\kappa$ as exactly constant.
Equation~\ref{eq:fluxhydrostat} then reduces to
\begin{equation}\label{eq:reducedhydrostat}
\frac{1}{r}{\partial \over \partial r}\left[ r^2\Omega^2 \left(1 - j^2\right)\right] +
            {\partial \over \partial z}\left[ z \Omega^2\right] = 0.
\end{equation}
Although the black hole mass may be larger than the stellar mass enclosed within
the torus, it is easy to allow for potentials more general than a simple point-mass
by writing $\partial\ln\Omega/\partial\ln r = -\alpha$.   Then, for any
particular potential described by $\Omega(r)$, equation~\ref{eq:reducedhydrostat}
determines the unique $j(r)$ permitting a self-consistent solution.   Written in
terms of $\alpha$, equation~\ref{eq:reducedhydrostat} becomes
\begin{equation}
r{dj^2 \over dr} + 2\left(1-\alpha\right) j^2 = 3 - 2\alpha,
\end{equation}
which has the solution
\begin{equation}\label{eq:jsqsoln}
j^2(r) = \left[j_{\rm in}^2 + f(\alpha)\right]\left(\rnorm\right)^{2(\alpha-1)}
              - f(\alpha),
\end{equation}
where $r_{\rm in}$ is the radius at which the higher-energy photons are converted to
infrared, $j_{\rm in} = j(r_{\rm in})$, and $f(\alpha) = 0.5(3-2\alpha)/(\alpha-1)$.
When $\alpha=3/2$ (a point-mass potential), $f(\alpha) = 0$ and
$j(r) = j_{\rm in}(r/r_{\rm in})^{1/2}$.  When $\alpha = 1$ (a logarithmic potential),
a logarithmic dependence replaces the power-law:
$j^2(r) = j_{\rm in}^2 + \ln(r/r_{\rm in})$.

\begin{figure}
\centerline{\includegraphics[angle=90,scale=0.8]{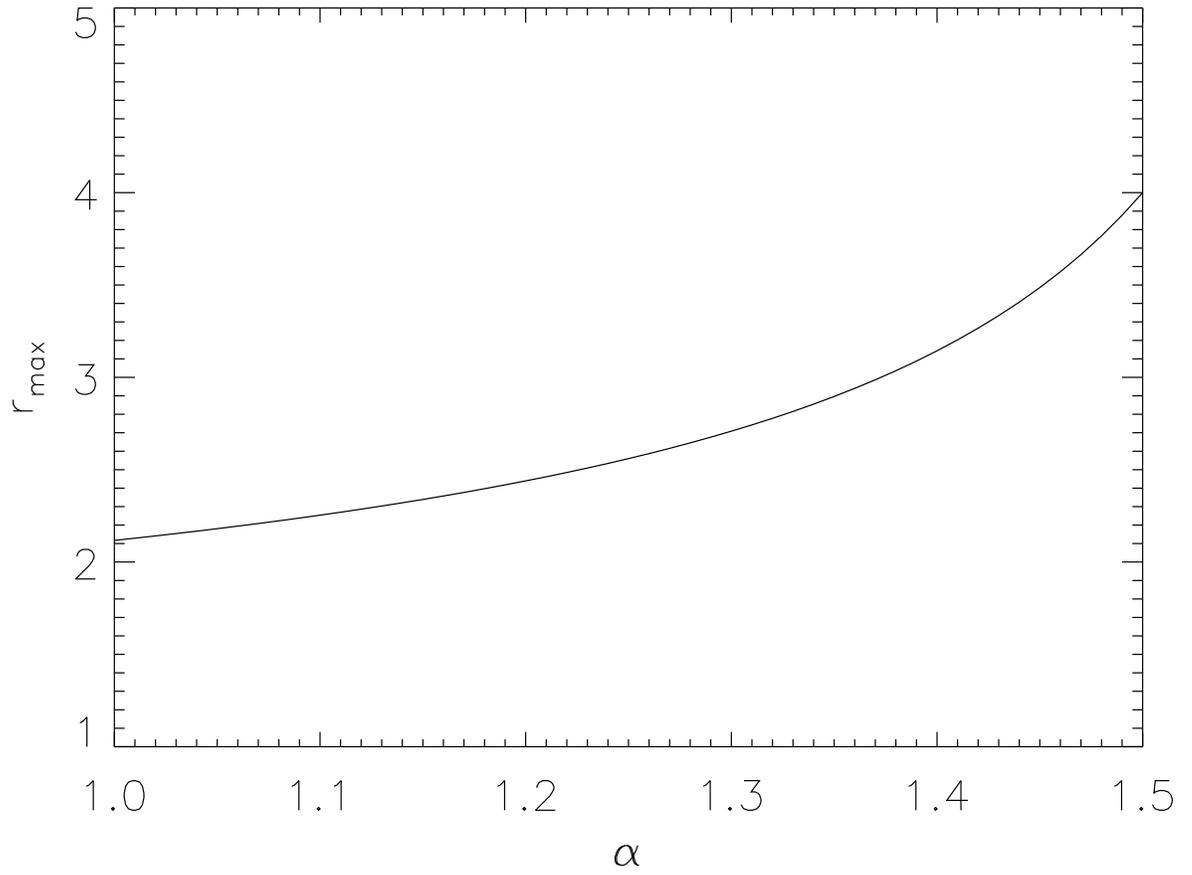}}
\caption{Maximum torus radius as a function of $\alpha$ for fixed $j_{\rm in}$.
The case illustrated is for $j_{\rm in} = 0.5$.
\label{fig:rmax}}
\end{figure}

     Assuming that the point where $j=1$ marks the outer edge of the torus, we can
use this solution to find the span of radii over which the torus exists.  For
$\alpha \neq 1$, it stretches from $r_{\rm in}$ to
\begin{equation}
r_{\rm max} = r_{\rm in}
   \left[\frac{1 + f(\alpha)}{j_{\rm in}^2 + f(\alpha)}\right]^{1/[2(\alpha-1)]};
\end{equation}
in the special case of the logarithmic potential, $r_{\rm max} = \exp(1-j_{\rm in}^2)$.
As Figure~\ref{fig:rmax} illustrates, for fixed $j_{\rm in}$, the breadth of these
tori stretches
moderately as the slope of the potential steepens: for $j_{\rm in} = 0.5$,
$r_{\rm max}$ rises from $\simeq 2.1 r_{\rm in}$ for $\alpha = 1$ to $4r_{\rm in}$ for
$\alpha = 1.5$.  Not surprisingly, $r_{\rm max}$ increases with diminishing
$j_{\rm in}$ at fixed $\alpha$.

     With $j^2(r)$ known, equation~\ref{eq:eraddyn} may be separated into two
equations for $\rho$, which must be consistent with each other:
\begin{equation}\label{eq:rhoconsist}
\rho = -\frac{1}{3z\Omega^2}\frac{\partial E}{\partial z} =
       -\frac{1}{3r\Omega^2\left[1-j^2(r)\right]}\frac{\partial E}{\partial r}.
\end{equation}
It is instructive to rewrite in characteristic form the partial differential
equation for $E$ implied by the second equality:
\begin{equation}
{dE \over ds} = \frac{\partial E}{\partial z}\frac{dz}{ds} + 
               \frac{\partial E}{\partial r}\frac{dr}{ds} = 0,
\end{equation}
with
\begin{equation}
{dz \over ds} = {1 \over z} \qquad {dr \over ds} = -{1 \over r\left[1-j^2(r)\right]}.
\end{equation}
From the characteristic form, we see that $E$ is constant along contours parameterized
by the pair of characteristic equations
\begin{equation}
{1 \over 2}z^2 = s, \qquad 
     \int_{r_*}^r \, dr^\prime \, r^\prime \left[1 - j^2(r^\prime)\right] =
                   -s + \lambda,
\end{equation}
where $r_*$ is arbitrary and we set the integration constant for $z(s)$ to zero by
choosing $z=0$ at $s=0$.   The radius at which a contour labelled by $\lambda$
passes through $z=0$ is given implicitly by the second characteristic equation.
Equating the two expressions for $s$ gives an explicit definition of the contours
of constant $E$:
\begin{equation}
\frac{1}{2}z^2 + \int_{r_*}^r \, dr^\prime \, r^\prime \left[1 - j^2(r^\prime)\right] 
              = \lambda.
\end{equation}
Evaluating the integral, we find:
\begin{equation}
\frac{1}{2}\left(\znorm\right)^2 + \frac{1}{4(\alpha-1)}\left(\rnorm\right)^2
-\frac{1}{2\alpha}\left[j_{\rm in}^2 + f(\alpha)\right]
     \left(\rnorm\right)^{2\alpha}  = \lambda,
\end{equation}
where we have redefined $\lambda$ so as to absorb a number of terms that depend on
the (arbitrary) $r_*$.

      Because $j(r)$ depends only on $\Omega(r)$, the shapes of these contours
depend only on the shape of the gravitational potential and on the boundary
condition $j_{\rm in}$.   Generically, they are closed curves with principal
axes parallel to the $r$ and $z$ axes, and elongated
in the $z$ direction.   For example, when $\alpha=3/2$, the contours are given by
\begin{equation}
\frac{1}{2}\left(\znorm\right)^2 + \frac{1}{2}\left(\rnorm\right)^2 - 
       \frac{1}{3}j_{\rm in}^2\left(\rnorm\right)^3 = \lambda.
\end{equation}
That is, the larger $j_{\rm in}$ is, the more the contours stretch in the $z$
direction.

    We may solve an ordinary differential equation for $E(\lambda)$ along any path
on which $\lambda$ varies monotonically.   For example, if we choose the path to run
outward along the $r$ axis from $r_{\rm in}$, we have
\begin{equation}
\frac{dE}{d\lambda} = \frac{\partial E}{\partial r} \frac{dr}{d\lambda} =
     -3 \rho \Omega^2.
\end{equation}
In other words, knowledge of $\rho(\lambda)$ on this path is a prerequisite for
accomplishing such a solution.  Physically, this should come as no surprise:
The run of radiation energy
density with position must certainly depend on how much matter there is with which
to interact and how it is distributed.  Somewhat arbitrarily, we choose to supply
$\rho(r,z=0)$ and integrate
\begin{equation}\label{eq:ode}
\frac{dE}{dr} = -3\rho(r,0)r\Omega^2 \left[1 - j^2(r)\right].
\end{equation}

    It is convenient in this context (in which we have already written $\Omega
\propto r^{-\alpha}$) to consider a density boundary condition that is also a
power-law in radius: $\rho(r,0) = \rho_{in}(r/r_{\rm in})^{-\gamma}$.
With this choice of $\rho(r,0)$, every term on the right-hand-side of
equation~\ref{eq:ode} is a power-law and the equation may be integrated exactly.
The result is
\begin{equation}
E(r) = E_{\rm in} - 3\rho_{\rm in}r_{\rm in}^2\Omega_{\rm in}^2
  \left\{\frac{1 + f(\alpha)}{2-2\alpha-\gamma}
  \left[\left(\rnorm\right)^{2-2\alpha-\gamma} - 1\right] + 
  \frac{j_{\rm in}^2 + f(\alpha)}{\gamma}
      \left[\left(\rnorm\right)^{-\gamma} -1\right]\right\}.
\end{equation}
Note that if the power-law for $\rho(r,0)$ were to extend to $r=\infty$,
securing a finite optical depth would require $\gamma > 1$.  However, because we
cut off the power-law at a finite maximum radius, this restriction on $\gamma$ is
eliminated.

    Once $E(r)$ in the equatorial plane has been found, one may directly
determine $E$ at {\it all} $r$ and $z$ by extending that solution into the
plane following the characteristic curves that define its constant-value
contours.  The density $\rho$ then follows from either of the
equations~\ref{eq:rhoconsist}.

\subsection{Governing parameters, boundary conditions, and range of validity}

     So far we have identified three parameters that govern the character of these
solutions: $\alpha$, $j_{\rm in}$, and $\gamma$.   There are two more.   One of
these is $\tau_* \equiv \kappa\rho_{in}r_{\rm in}$, which sets the optical depth
scale.  If the density declines outward, it must be at least several, or there
will be nowhere where the diffusion approximation is valid.
The other parameter is $Q \equiv 3\rho_{in}r_{\rm in}^2\Omega^2(r_{\rm in})/
E(r_{\rm in},0)$, whose physical
meaning is most clearly seen when rewritten in terms of more familiar quantities:
\begin{equation}
Q = 3\frac{\tau_*}{h}\frac{M(<r_{\rm in})}{M_{BH}}\frac{\kappa_T}{\kappa}
     \frac{L_E}{L},
\end{equation}
where $M(<r_{\rm in})$ is the total mass interior to $r_{\rm in}$, and $M_{BH}$ is
the mass of the central black hole alone (which is what determines $L_E$ in the
usual definition).  The factor $h$ is the amount by which the optical depth of the
torus enhances $E(r_{\rm in},0)$ over the value it would have if the radiation could
stream out freely.   Because we expect the torus to be optically thick in the
mid-infrared, but radiation can escape freely through the axial hole,
$h \sim \tau_*/[1 + (\tau_* - 1)\phi]$, where $\phi$ is the fraction of solid
angle that is open as viewed from the position $(r_{\rm in},0)$.  From the statistics
of type~1 versus type~2 AGN, we might suppose that $\phi \lesssim 1/2$.   If
$\tau_* \sim 10$--30, then $h \gtrsim 2$.  Reasonable values of $Q$ would then be
$\sim 0.3$--30.

     Our approximations do not hold at all locations, so it is important to mark
off carefully the region of the $r$-$z$ plane in which we can apply them.
Because the phenomenology points us strongly toward a density distribution
with an inner hole, we need to define an inner radial edge, within which
the matter is optically thin.   Unfortunately, we do not know the shape
of this inner edge {\it a priori}; indeed, that shape is likely the result
of some rather complicated dynamics, as will be discussed in the next
several paragraphs.   In this paper, whose goal is merely to demonstrate
the possibility of thick structures supported by radiation pressure, we simply
choose a vertical inner edge, i.e., $r_{\rm edge}(z) = r_{\rm in}$.
Even if $r_{\rm edge}$ were a function of $z$, the solution we have
derived in the previous subsection would still hold wherever $r > r_{\rm edge}$.
The extension of $E(r,0)$ into the full $r$--$z$ plane is entirely independent
of the exterior boundary conditions because the hydrostatic balance condition
substitutes, in effect, for boundary conditions in determining
the interior diffusion solution.  Reconciliation of our interior
solution with the exterior boundary conditions will also be discussed later in
this section.  When we know more about $r_{\rm edge}(z)$,
that knowledge will change the size and shape of the region where our
hydrostatic diffusion solution applies, but will not not alter its
nature in the torus interior. 

     Unfortunately, a proper determination of the position of the
inner edge is far beyond the scope of this initial effort (Pier
\& Voit 1995 present a simplified model).  Although
the problem is superficially similar to the one treated here (simultaneously
solving the equations of 2-d radiation transfer and 2-d hydrostatic equilibrium),
several new effects become important at the edge and drastically complicate
the problem.  There is a large outward force from absorption
and scattering of the optical/UV continuum arriving directly from the nucleus;
there is also a comparably large, but inward, force due to the infrared flux emerging
from the torus.  Because the most important effects for both occur across
photospheres (at different places for the different wavelengths, of course)
the diffusion approximation is wholly inadequate for both.
Gas pressure gradients, although (by assumption in this model) unimportant
in the bulk of the torus, can also become significant locally.

In one respect, evaluating the force exerted by the optical/UV continuum
is relatively simple: it has a single point-like source at the nucleus, so
its transfer problem is essentially one-dimensional along radial rays
(departures from true one-dimensionality arise only to the degree that
the albedo of dust permits scattering to spread the beam).
On the other hand, the opacity, far from being roughly
constant, is a strong function of the radiation intensity, making this
transfer problem highly nonlinear.  At the inner edge of the torus,
the gas can be photoionized and, in some cases, radiative heating
can warm the dust above its sublimation temperature.  In addition,
photoionization heating may raise the gas's temperature high enough
for it to destroy the dust by sputtering. 

The only simple aspect of the infrared transfer problem is its qualitative
behavior in two extreme geometric limits: In the limit of infinitesimal opening
angle, the infrared intensity would be identically constant across the central
hole; in the limit of very large opening angle, it would be considerably
smaller along the axis than at the torus's inner edge.  However, because its principal
opacity is also due to dust, many of the same difficulties that apply to the
optical/UV continuum at the torus's inner edge also apply to the infrared.
In addition, because its sources are distributed, the solution is thoroughly
2-d and global.  At the order of magnitude level, the infrared intensity
in the central hole is likely to be
comparable to or greater than the intensity of the optical/UV
continuum precisely because of the blanketing provided by the optically
thick torus; we have hidden this ratio in the $h(\tau_*)$ fudge factor
already introduced.  The total force it exerts on the gas in the inner
edge region is determined by the contrast in intensity between the
main body of the torus and the axis.  In rough terms, we expect this
fractional contrast to increase from the midplane upward because the
fraction of ``open sky" seen from a position on the axis increases upward,
but to be quantitative about this dependences requires a proper 2-d global
transfer solution.

Further complications can be caused by strong local gas pressure gradients,
which are entirely absent in our model for the torus interior.  Because
the opacity of dust in the optical/UV is considerably
greater than in the infrared, the radiation force due to this band is
expressed across a much narrower zone than that due to the infrared.
As an immediate consequence, sharp gas pressure gradients are likely
to be created across these short lengthscales, even though the total gas
pressure contrast from the torus body to the outside
may not be that large.  Gas pressure effects introduce
further nonlinearity into the transfer problem because the gas's
equation of state is also strongly dependent on the radiation
intensity. 

Summarizing this qualitative discussion, we expect the location and shape of the
inner edge to be the immediate result of balancing the opposing radiation and gas
pressure gradient forces in the context of a rapidly changing physical
state for the gas.  Over longer timescales, mass flux balance will also come
into play.  As matter is ionized and heated at the extreme inner edge of the
torus, it rushes away toward the lower pressures found at higher altitudes
in the torus hole and beyond (Krolik \& Begelman 1986, Balsara \& Krolik 1993).
To achieve a steady state, matter must accrete through the torus in order to
replace the evaporated matter.  The accretion rate is controlled
by angular momentum transport; if, as is common in other disks, this is due
to MHD turbulence, the problem is further complicated.  Thus, determining
the position of the inner edge is a much more difficult problem than solving for
the static structure of the torus interior.

The outer radial edge is determined by the requirement that $j(r) \leq 1$; greater
$j$ would make hydrostatic equilibrium impossible.  Although there is no automatic
physical inconsistency created by placing the outer radial edge where $j < 1$,
doing so begs the question of the dynamical state of the matter beyond: how
does it make the transition from partial radiation force support to rotational
support?  The diffusion approximation is valid
only within those regions whose optical depth to infinity is $> 1$.  Consequently,
our solution should not be extended beyond the surface on which the vertical optical
depth $\tau_z = \int_z^\infty \, dz^\prime \, \kappa\rho(r,z^\prime) = 1$.
Even if our approximations remained valid at small optical depth, one might define
the torus edge as its infrared photospheric surface in any case.

     At the photosphere we have another boundary condition, but one that can
be applied only approximately: the flux as estimated by the diffusion approximation
should roughly match the flux as estimated on the basis of free-streaming.
Because we equate the diffusive flux with the flux necessary for hydrostatic
balance, this condition amounts to requiring that
\begin{equation}\label{eq:opthinbc}
|\vec g_{\rm eff}/\kappa| \sim E(r,z)
\end{equation}
where $\tau_z = 1$.  Unfortunately, the only way we can locate the
photosphere is in terms of the density distribution derived from the diffusion
equation.  Outside the photosphere, this density distribution cannot be
completely correct, yet what we mean by the location of the photosphere is
the curve $z_{ph}(r)$ defined by $\int_{z_{ph}}^{\infty} dz \rho(r,z)\kappa = 1$.
In other words, we find the photosphere using the density distribution in
exactly that region where we know it least well.  A further uncertainty
is introduced by the fact that we can estimate to an accuracy of only
a factor of $\sim 3$ the radiation energy density required to carry the flux
in the optically thin regime.  For all these reasons, we
require equation~\ref{eq:opthinbc} to be satisfied only to within a factor of 3.

     There is also one additional constraint on acceptable solutions: if at anywhere
along the radial axis $E<0$, the solution is obviously unphysical.  In practise,
we find that the photospheric boundary condition is best matched at the smallest
$\gamma$ such that $E(r) > 0$ everywhere in the range
$r_{\rm in} \leq r \leq r_{\rm max}$.

\section{Results}

\begin{figure}
\centerline{\includegraphics[angle=90,scale=0.4]{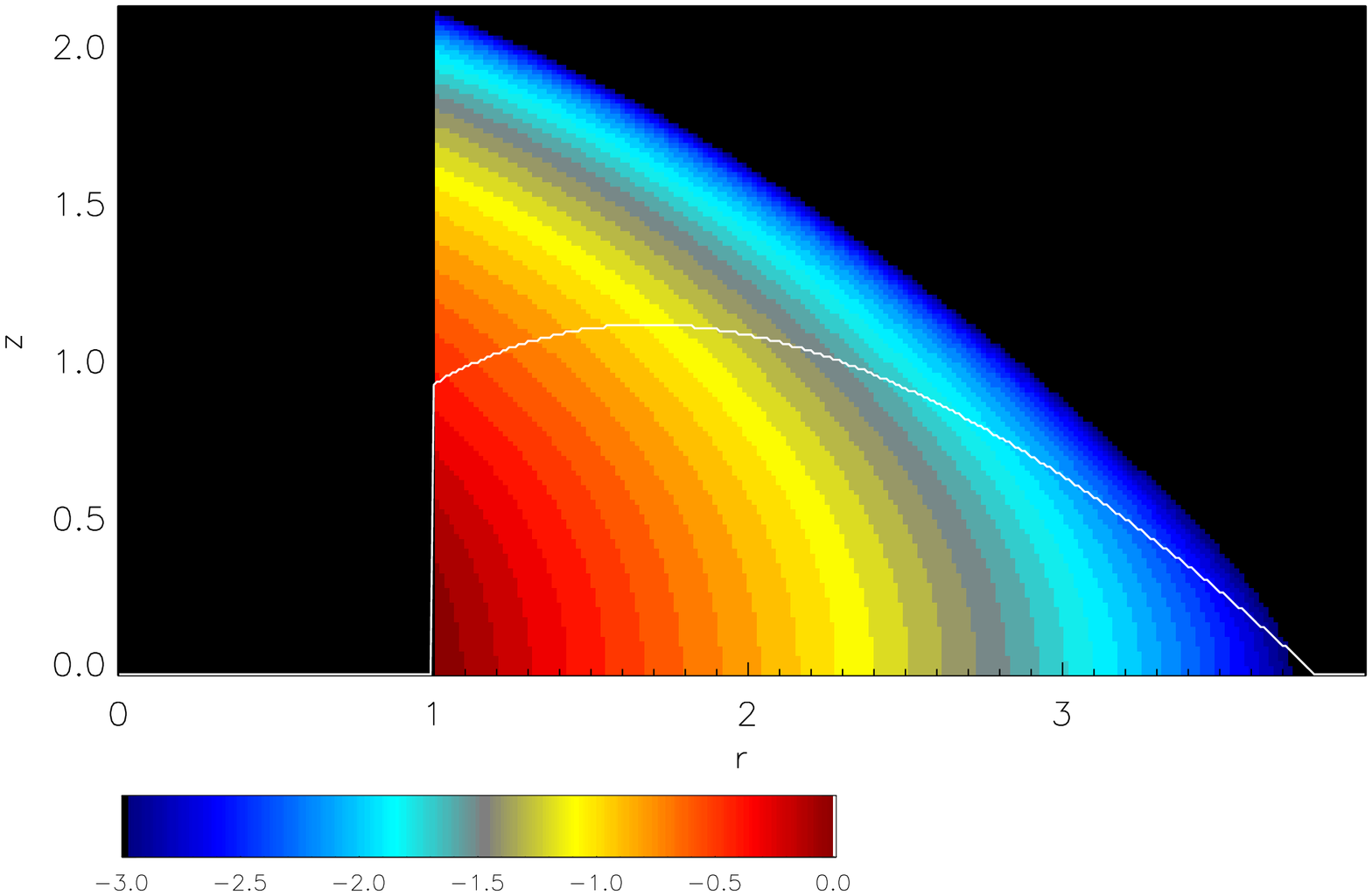}
            \includegraphics[angle=90,scale=0.4]{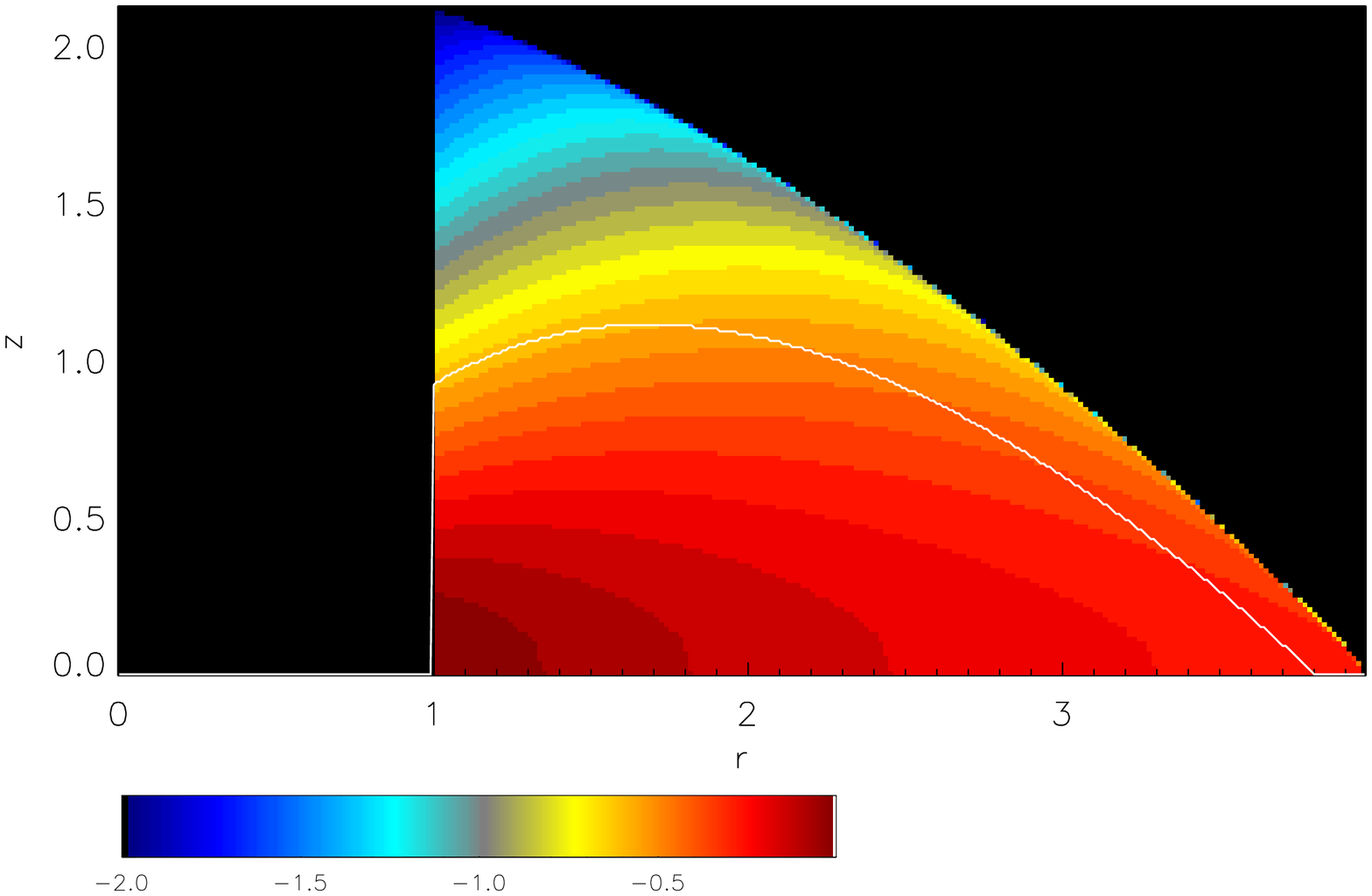}}
\caption{A solution with $j_{\rm in} = 0.5$, $\alpha = 1.5$, $\gamma=0.5$, $\tau_*=10$,
and $Q=3$.  Left panel: Radiation energy density.  Right panel: Matter density.
In both, the scale is logarithmic, and the white curve shows the surface on which
$\tau_z = 1$, the photosphere on the top of the torus.
\label{fig:typresult}}
\end{figure}

    With these thoughts in mind, consider the ``typical" parameters
$j_{\rm in} = 0.5$, $\alpha = 1.5$, $\tau_*=10$, and $Q=3$.  That is,
rotational support is substantially depressed at the inner edge, the potential
is that of a pure point-mass, the column density in the midplane is
$\sim 10^{24}$~cm$^{-2}$, and $h L/L_E \simeq 1/3$--1.
An acceptable solution requires $\gamma \simeq 0.5$, that is, the density declines
slowly outward in the equatorial plane.  If the matter density declines more
steeply, there is too little optical depth in the torus, and the energy density
outside the photosphere predicted by the diffusion approximation
is substantially larger than what is required to carry the flux in the optically
thin regime; if the matter density declines more slowly, the optical
depth is too great, forcing $E$ to go negative.  The successful solution that
results from a compromise between these two extremes (i.e., $\gamma = 0.5$)
is illustrated in Figure~\ref{fig:typresult}.  Full Keplerian support is reached
at $r_{\rm max} = 4r_{\rm in}$.  As expected, the contours of radiation
energy density inside the torus are extended upward; this is exactly what one
would expect when the radiation finds it easier to move vertically than radially.
Note that, by assumption, the local temperature $T = (E/a)^{1/4}$, so
the greatest temperature is found at $(r_{\rm in},0)$, and it declines
upward and outward from there.  Contours of constant density, on the other hand,
are extended radially.  This, too, is entirely in line with expectations, given
the difficulty of vertical support against gravity.

    In both panels of Figure~\ref{fig:typresult}, a white curve shows the location
of the photospheric surface on the top of the torus.  Formally, our solution is
invalid outside this white curve, as the diffusion approximation does not well
describe the relation between energy density and flux in optically thin regions.
Although much of the volume shown is in the optically thin region, most of its
mass is within the optically thick portion of the torus.  Consequently, while
not taking too seriously the details of the solution in the optically thin zone,
we can also be assured that they will not have serious impact on the issue
of greatest concern here: the mass distribution within the torus.

     Having located the outer boundaries within which this solution is physical,
it is now time to consider its behavior on the inner boundary.  As discussed
earlier, we are not prepared to present a proper global infrared transfer
solution that includes the central axial hole.  Qualitatively, however,
we might expect that the radiation intensity inside the torus hole would fall
with height somewhat faster than $\propto (r_{\rm in}^2 + z^2)^{-1}$ because
the confinement due to the torus walls diminishes as its surface is neared.  In
fact, in our fiducial model, $(r_{\rm in}^2 + z^2)E(r_{\rm in},z)/E(r_{\rm in},0)$
falls slowly with increasing $z$, reaching $\simeq 0.6$ where the photosphere
intersects the inner edge.  This sort of behavior therefore appears to be
at least loosely consistent with what one might guess about transfer solutions
inside the torus hole.

     The density profile in the equatorial plane is a strong function of $Q$ when
all other parameters are held fixed.  Larger $Q$ demands a steeper profile: as it
changes over the factor of 10 from 0.6 to 3.0 to 6.0, the density power-law demanded
rises from -2.58 to 0.50 to 3.05.  In other words, the density must fall more
steeply outward when $L/L_E$ is smaller.  When $\gamma < 1$, of course,
the integrated
optical depth is dominated by the outermost radius; solutions in this regime
must involve very sharp cut-offs in the density profile at the outer radius.

      Greater rotational support at the inner edge means that full Keplerian
angular momentum is reached at a smaller radius.  For example, if
$j_{\rm in} = 0.75$ but the other parameters are held fixed at their fiducial
values, $r_{\rm max} = 1.76r_{\rm in}$, and the matter
density must rise outward quite steeply ($\gamma = -4.71$) for there to be
enough optical depth across this diminished radial thickness; otherwise,
the boundary condition matching the diffusion approximation energy density
and the free-streaming energy density cannot be satisfied.  Conversely,
if the fractional angular momentum support at the inner edge is much
smaller, $r_{\rm max}$ is much larger and the density
falls steeply outward: for $j_{\rm in} = 0.25$, $r_{\rm max} = 15.8r_{\rm in}$
and $\gamma = 1.71$.

       It is possible that distributed stellar mass may contribute significantly
to the gravitational potential in the torus region.   If so, orbital speeds
will decline more slowly with increasing radius than in the Keplerian prediction.
When this is the case, the equatorial density profile able to produce
an equilibrium switches from one that gradually falls outward to one that
gradually rises.  For example, for $j_{\rm in} = 0.5$
and $Q=3$, $\gamma$ drops from 0.5 to -0.19 as $\alpha$ falls from 1.5 to 1.1.

       The optical depth parameter $\tau_*$ does not appear explicitly in the
differential equations, so the shapes of the matter and energy density contours,
as well as the $\gamma$ required by the radiation energy density
boundary condition at the photosphere, do not depend on it.  However, the
outline of the torus that results {\it does} change with $\tau_*$ because
the photosphere (not surprisingly) rises higher and higher with increasing
optical depth.

\section{Consequences and Comparison with Observations}

       We have found that hydrostatic radiation-supported solutions can be
found for plausible parameters, particularly when $Q$ is within a factor of
several of unity.  For these solutions to be found in real AGN, two additional
conditions must be satisfied: the matter's angular momentum must be
redistributed so that the equilibrium $j(r)$ is achieved; and the radial
matter density profile in the equatorial plane must be adjusted to the
right shape.   It is possible that both may be achieved as a result of
the magneto-rotational instability and the magnetic torques created by
the MHD turbulence it drives.  This is because $j_{\rm in}$ may be
viewed as a function of $\gamma$: a matter density profile in the
equatorial plane that does not fall as rapidly, or even rises outward
(i.e., a smaller $\gamma$) is consistent with dynamical
equilibrium when the fractional rotational support at the inner edge is greater.

     Whether this actually happens in practise is uncertain.  In
conventional global disk simulations, MHD turbulence very efficiently redistributes
angular momentum until it reaches a nearly-Keplerian radial profile
(De Villiers et~al. 2003).  If that happened here, hydrostatic balance
would be possible only if accompanied by a substantial positive
$\nabla \cdot \vec {\cal F}$.  This might perhaps be supplied by local
sources of heat, such as stars or Compton-heating due to hard X-rays
that penetrate the bulk of the torus, but if sources such as these
are inadequate, the disk
would be forced to collapse to a thin configuration.   On the other
hand, in this context the equilibrium angular momentum distribution
reached as a result of MHD turbulence might actually be the one required
by this model (eqn.~\ref{eq:jsqsoln}) because the net ``gravity" has
been effectively reduced by the outward radiation force.

    In this context, it is notable that the velocity profile of the maser
spots in NGC~1068 is $\propto r^{-0.3}$, rather than $\propto r^{-0.5}$
as one might expect from circular orbits in a point-mass potential (Greenhill
et~al. 2006).   Our model predicts a rotational speed that always declines
more slowly outward than simple circular orbits in the gravitational potential
would dictate; for example, the mean rotational speed is constant
as a function of radius when $\alpha = 1.5$.  Thus, a shallow rotation
curve may not signal a stellar contribution to the gravitational
potential---it could instead be a symptom of radiation support.

    Maser kinematics can also be used to estimate the central mass, either
from the magnitude of the circular speeds or from the acceleration $a$ seen
in maser emission on the direct line of sight to the nucleus (although
the circular speed is a more easily-observed quantity, $a$ has been
measured in four examples: Henkel et al. 2002 and references therein).  In the
former case, the mass inferred on the basis of Keplerian orbits is $v^2 r/G$,
in the latter, $a r^2/G$.  The sub-Keplerian rotation that directly
follows from the presence of radial radiation forces means that this
inferred mass is an underestimate of the true mass by a factor $j^2$
in both cases.  If the maser emission is driven by X-ray excitation
(Neufeld et~al. 1994), it takes place very close to $r_{\rm in}$, so the
relevant value of $j$ is $j_{\rm in}$.

     Interestingly, the shape of the angular momentum profile required
for equilibrium depends only on the underlying gravitational potential,
and in this respect is a comparatively robust prediction of the model.
On the other hand, because $j_{\rm in}$ is related to $\gamma$ through
the detailed equilibrium solution, which in turn depends on approximations
like the photospheric boundary condition, their quantitative relationship
is dependent on the quality of the several approximations made here.

       We have also seen that a relatively small change in $Q$ requires
a large change in density profile: $\gamma$ falls from $\simeq 3$ to
$\simeq -2.5$ when $Q$ falls from 6 to 0.6.  If the optical depth is
held fixed, $Q$ is primarily dependent on $L/L_E$, to which it is
inversely proportional.   A consequence of this model, therefore, is
that geometrically thick tori may be associated only with AGN having
$L/L_E$ within a range not much greater than a factor of 10, logarithmically
centered on $L/L_E \sim 0.3$.  Although the measurement of Eddington
luminosity ratios is still very difficult, this range is easily
consistent with the data in hand (e.g., the maser-based measurement of
the black hole mass in NGC~1068: Greenhill et~al. 1996, Gallimore et~al. 1996;
the somewhat shakier black hole masses from reverberation-mapping:
Metzroth et~al. 2006 and references therein; or the still shakier inferences
from photoionization-scaling: McLure \& Dunlop 2004).

       More precisely, if stellar contributions to the gravitational
potential are negligible, the fact that $\kappa_T/\kappa \sim 0.03$--$0.1$ for
warm dust implies that tori should be puffed up by radiation whenever
$L/L_E \sim (0.03$--$0.1) \tau_*/h$.   The ratio $\tau_*/h \sim 1 + (\tau_* - 1)\phi$,
so it may vary over a modest range, from $\sim 1$ to a few, depending on
the total optical depth through the torus and the shape of the inner hole.
Infrared radiation pressure cannot provide
predominantly vertical support against gravity unless the
torus is optically thick in the mid-infrared; if it were optically thin,
the flux would emerge more or less radially.  When $\gamma > 1$, this
requirement means that $\tau_*$ cannot be less than a few.  However, when
$\gamma < 1$, as is often the case, most of the optical depth is found
near the outer edge of the torus, so the lower limit on $\tau_*$ can
be smaller.  At the low end of the permitted optical depth range, we
would also expect $h$ to be no more than a few.  Larger optical
depth cannot lead to much greater values of $h$, however, because
the axial hole through the torus creates an escape channel.  This
is why, although solutions can be found for arbitrarily large values of
$\tau_*$, they are unrealistic unless somehow $L/L_E$ can be $\gg 1$.

      Measurements of the total optical depth in the equatorial plane
of obscuring tori are difficult to come by.  The column density
of matter on the line of sight can be measured directly by soft X-ray
absorption (or at least a lower bound placed when the Compton depth
is greater than unity).  However, in most cases, we do not have any
direct evidence of the inclination angle of the torus to our line of
sight; even if we did, it would be difficult to constrain directly
the optical depth along the equatorial plane, in general a direction
oblique to our line of sight.

\begin{figure}
\centerline{\includegraphics[angle=90,scale=0.8]{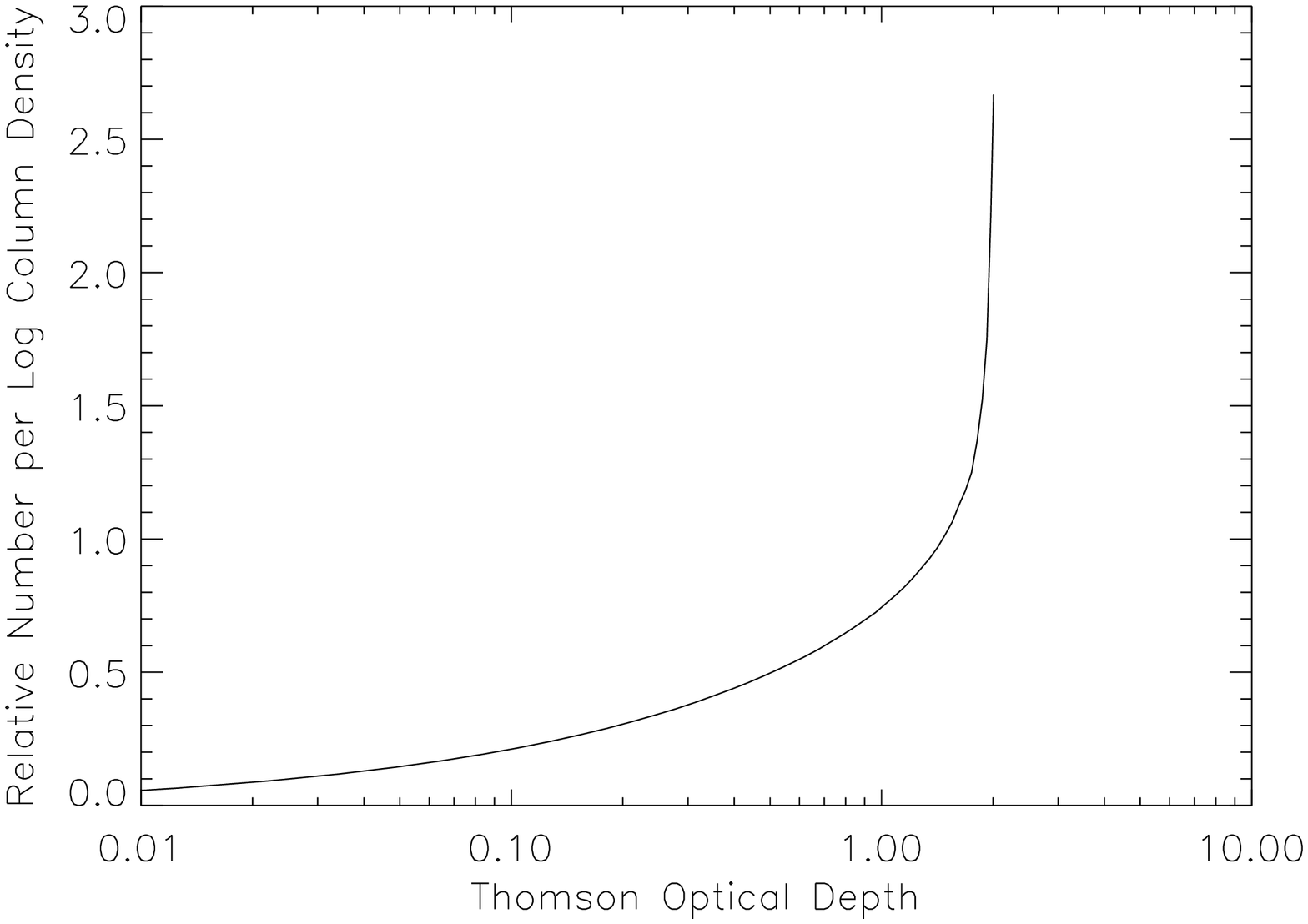}}
\caption{The predicted column density distribution for the solution with
the same parameters as Fig.~\ref{fig:typresult}.
\label{fig:coldensdist}}
\end{figure}

      On the other hand, we can measure the statistical distribution
of column densities to obscured AGN and compare it to the predictions
of this model, although this calculation is somewhat sensitive to the
looseness in our location of the torus inner edge.  The predicted probability of
seeing a given column density is simply proportional to the solid angle
associated with the polar angle producing that column.  Typically,
these solutions predict
a wide range of associated column densities because lines of sight farther
from the equatorial plane pass through densities exponentially lower than
those closer to the plane.   Because we are primarily interested in $\tau_*
\sim 10$, which corresponds to $\tau_T \sim 0.3$--$1$, the range of expected
Thomson depths is from $\sim 10^{-2}$ to $\gtrsim 1$, in good correspondence
with observations (Risaliti et~al. 1999, Treister et~al. 2004).  However,
in contrast to the observational results, which tend to show a flatter
distribution, there is a tendency for most of solid angle to be associated
with the higher column densities.  For example, if we take $\kappa_T/\kappa = 0.1$,
in our fiducial case the number of systems per logarithm of column density with
$\tau_T \simeq 1$ is predicted to be $\sim 10 \times$ the number seen with
$\tau_T \simeq 0.01$ (Fig.~\ref{fig:coldensdist}).
Qualitatively, the shape of this distribution has two sources:
the solid angle per unit polar angle is, of course, greatest
near the equatorial plane, where the densities are also greatest; and
the high density region stretches away from the equatorial plane because
the vertical component of gravity is relatively small there ($g_z \propto z$).
Given the approximate character of our model, we expect that the prediction of
a broad range of observed column densities should be fairly robust, but
the exact shape of the predicted distribution is subject to significant uncertainty.

    A different comparison with observations relates to the most fundamental
reason why we believe the obscuration is geometrically thick: if the number of
obscured AGN is comparable to or greater than the number of unobscured AGN having
the same bolometric luminosity, then much of the sky surrounding the nucleus
must be optically thick in the optical and ultraviolet.   In principle, the
solutions we have described also predict the fraction of solid angle
obscured by the torus.  For this question, however, the arbitrariness of our
inner edge introduces an especially large systematic uncertainty.  Because only
small column densities are required to stop UV photons if there is a normal
dust/gas ratio, any curvature of the inner edge could be significant.  The following
numbers should therefore be taken more in the way of examples than as
serious predictions.  Nonetheless,
if we temporarily leave aside these considerations, we can still estimate
the obscured solid angle by taking these models at face-value.  If the
wavelength of interest has an opacity $100\kappa$
(corresponding to $\lambda \simeq 4000(10\kappa_T/\kappa)$~\AA:
Draine \& Lee 1984), $89\%$
of solid angle is opaque in our fiducial model.  This corresponds to a
half-opening angle of $27^\circ$.  Not surprisingly, larger $Q$ (lower
luminosity relative to Eddington) leads to a smaller obscured fraction,
but only slightly smaller: $Q=6$ (for all other parameters fixed)
makes a torus that blocks $85\%$ of the sky ($32^\circ$ half-opening
angle).

      In summary, we have constructed a simple, and entirely analytic,
model of how infrared radiation pressure can, in principle, support
obscuring tori around AGN.  The model employs numerous approximations
and simplifications that would undoubtedly be improved in a more complete
and realistic picture.   Most notably, a true transfer solution,
rather than one adopting the diffusion approximation, would allow
a proper connection to the boundary conditions at the photosphere of
the obscuring matter.  Frequency-dependent opacities would also improve
its realism, although probably not as dramatically.  In its treatment of
the gas, a more realistic model might allow for clumping, both in regard
to the opacity and to permit the introduction of supersonic random motions.
However, we also wish to point out that, to the degree vertical support
is provided by radiation pressure, the necessity of supersonic motions---and
therefore clumping---is diminished.

      This model also raises a number of questions.  For example, as
already discussed, the required angular momentum profile may or may not
be achieved.  It is also unclear whether this equilibrium is stable to
a variety of perturbations---smooth motions in the gas, clumping in the gas,
departures from the equilibrium angular momentum profile---to name a
few possibilities.  It is similarly uncertain whether the equilibrium,
even if stable, can be reached from a wide range of initial conditions.

     Despite these questions, the complete analytic solvability of this
model means that we can learn from it a number of interesting
qualitative facts about radiation forces in this context.
Specifically, we have shown that a density distribution for the obscuration
can be found in which {\it both} radiation diffusion and dynamics are
in equilibrium.  This distribution, as demanded by the phenomenology of
AGN, can obscure a sizable solid angle for reasonable parameters.  We have
also found that in order for the equilibrium to be possible, several
conditions must be met: the luminosity of the nucleus must
be within a factor of several of $0.1L_E$; because there is always
a significant radial radiation force, the matter in the torus must
orbit more slowly than in a Keplerian orbit; the Thomson depth of
matter in the equatorial plane must be not too far from $\sim O(1)$;
and the level of rotational support relative to Keplerian is linked
to the radial profile of matter density in the equatorial plane, $L/L_E$,
and the total optical depth of the matter.

\acknowledgments{I would like to thank Eliot Quataert, Norm Murray, Phil
Chang, Todd Thompson, and Lincoln Greenhill for very stimulating conversations
and several very helpful suggestions.  I am also indebted to Eliot Quataert
and Omer Blaes for a careful reading of the manuscript, and to the second
referee, Mitch Begelman, for a thoughtful review.
This work was partially supported by NSF Grant AST-0507455 and NASA Grant
NNG06GI68G.}

\end{document}